\documentclass[12pt]{article}
\usepackage{epsfig}
\usepackage{graphicx}
\usepackage{amsmath}
\usepackage{amssymb}

\def\Title#1{\begin{center} {\Large {\bf #1} } \end{center}}
\usepackage{fancyhdr}
\begin{document}
\topskip 2cm

\Title{QCD at work: from lepton to hadron colliders and back}
\bigskip

\begin{raggedright}  

{\it \underline{Andrea Banfi}  \index{}
\footnote{Presented at Linear Collider 2011: Understanding QCD at Linear Colliders  in searching for old and new physics, 12-16 September 2011, ECT*, Trento, Italy}\\
Department of Physics, University of Freiburg, Germany\\
{\rm  andrea.banfi@physik.uni-freiburg.de}
}

\bigskip\bigskip
\end{raggedright}
\vskip 0.5  cm
\begin{raggedright} {\bf Abstract} The astounding Physics results
  obtained with high-energy colliders in the last two decades owe much
  to an impressive progress in the understanding of the
  dynamics of strong interactions. I give here a personal overview of
  how the advance in QCD triggered by the Physics of hadronic final
  states at LEP has been exploited for New Physics searches at the
  LHC. Conversely, the need for precision calculations for LHC
  experiments has stimulated a huge progress in the understanding of
  the all-order structure of gauge theories. These results raise high
  expectations on the status of QCD at the start of a linear
  collider.
\end{raggedright}

\section{Introduction}
\label{sec:intro}

With the start of the LHC Particle Physics has entered a new era. Not
only will we probably have a final answer on the mechanism of
spontaneous breaking of electroweak symmetry, but we could also
observe novel phenomena like dark-matter or black-hole production. The
LHC, being a hadron collider, can access a wide range of scales for
Physics beyond the (known) Standard Model, from the LEP boundary of
about 100 GeV up to the TeV scale. The price we have to pay is that
events appear contaminated by the presence of a large number of
hadrons. Among those, only a small fraction is related to the
short-distance processes we are interested in, the rest comes either
from secondary collisions of the remnants of the two broken protons
(underlying event), or even from further soft collisions occurring
within the same proton bunch, the so-called pile up. This is in sharp
contrast with the situation at LEP, where only a few tens of hadrons
were produced.
A further important difference between $e^+e^-$ and hadronic colliders
is the possibility of detection of individual hadrons. In the $e^+e^-$
case, the tracker has basically full solid-angle acceptance, so that
information on charged hadrons is available everywhere in rapidity.
At hadron colliders the tracker extends only inside a central region
spanning a few units in rapidity, whilst outside the only available
information comes from the calorimetric towers. Therefore, although
both LHC experiments are able to combine detector information into
objects like ``topo-clusters''~\cite{Aad:2009wy} or ``particle
flows''~\cite{Ball:2007zza} which should be quite close to individual
particles, at the moment the preferred objects for studies of hadronic
final states are just jets. This experimental issue has also
theoretical implications, as we will describe in the following.

QCD at LEP is a ``theory of hadrons'', whose dynamics can be
investigated from high momentum scales where quarks and gluons are
produced, to the low momentum scales where the hadronisation mechanism
is effective. This is done through final-state observables, like
event-shape distributions or jet rates, which combine in various ways
hadron momenta in numbers that provide an insight on the jet structure
or the geometry of each event. The typical situation at LEP is that
the hard scale of the process, the centre-of-mass energy $Q$, is much
larger than the hadronisation scale $Q_0$, which is of the order of
the mass of the proton. It is therefore possible to find values $Q_V$
for final-state variables such that $Q_0 \ll Q_V$, so that the
corresponding distributions can be reliably computed in perturbative
QCD. Furthermore, due to the fact that most $e^+e^-$ observables are
global, i.e.\ sensitive to emissions everywhere in the phase space,
and that in $e^+e^-$ annihilation one can safely rely on QCD
coherence, it is always possible to approximate multiple
soft-collinear parton matrix elements with a probabilistic angular
ordered branching~\cite{Banfi:2004yd}. This feature is the key of the
success in the description of QCD final states in $e^+e^-$
annihilation with both Monte Carlo event generators and analytical
calculations (for a review see~\cite{Dasgupta:2003iq}, and references
therein). Just to recall the impressive accuracy reached by QCD
calculations for hadronic final states in $e^+ e^-$ annihilation, a
fully differential code for $e^+ e^-$ into three jets is available to
order $\alpha_s^3$ (next-to-next-to leading order,
NNLO)~\cite{GehrmannDeRidder:2007hr,Weinzierl:2009ms}, all-order
resummation of large logarithms has been computed for selected event
shapes (thrust, heavy-jet mass) at next-to-next-to leading logarithmic
accuracy
(NNLL)~\cite{Becher:2008cf,Chien:2010kc,Monni:2011gb},\footnote{For
  the special case of the thrust even NNNLL accuracy is claimed in
  Refs.~\cite{Becher:2008cf,Abbate:2010xh}} and there exist also QCD
inspired analytical models for (leading) hadronisation
corrections~\cite{Abbate:2010xh,Dokshitzer:1995qm,Korchemsky:1999kt,Dokshitzer:1998pt}.

At the LHC QCD has to be the ``theory of jets'', since resolving
single hadrons is in general a difficult task. Although jet cross
sections are generally within the domain of perturbative QCD, there
are a number of features that make an all-order perturbative
description of jet observables problematic. First of all, jets
themselves are non-inclusive objects, there is no closed mathematical
expression that relates the momentum of a jet to the momenta of
final-state hadrons, not even approximately as happens for instance
for the thrust in $e^+e^-$ in the two-jet limit.  This makes it
impossible to write jet cross sections in terms of operator matrix
elements, as is done for many inclusive
observables~\cite{Collins:1984kg,Sterman:1986aj,Becher:2010tm,Ahrens:2008nc},
and sometimes also for event-shape
distributions~\cite{Becher:2008cf,Chien:2010kc,Stewart:2009yx}. Another
traditional worry expressed by all-order QCD practitioners is that jet
cross sections are generally non-global
observables~\cite{Dasgupta:2001sh,Dasgupta:2002bw,Banfi:2002hw}.
Non-globalness, together with the fact that the presence of two
hadrons in the initial state might spoil collinear
factorisation~\cite{Forshaw:2006fk,Catani:2011st}, cast serious doubts
on the applicability of coherent branching to jet observables in
hadronic collisions. At hadron colliders, especially at the LHC, there
is also a major concern about the separation of scales between
perturbative and non-perturbative Physics. Poorly understood phenomena
like underlying event of pile-up can add several tens of GeV's of
extra transverse momentum to QCD jets, causing huge distortions in
many commonly studied hadronic final-state observables (e.g.\
event-shape distributions)~\cite{Banfi:2010xy}.

Given this situation, it might seem that the knowledge of QCD we have
inherited from LEP is of little use for LHC Physics. While this
consideration might be true for precision studies (e.g.\ measurements
of $\alpha_s$), the insight on hadron dynamics we have at present can
be largely exploited for LHC phenomenology. This will be the subject
of the first part of my contribution (Section \ref{sec:LEP2LHC}). In
the second part (Section \ref{sec:LHC2LC}) I will shortly review the
progress in QCD triggered by the quest for precision calculations in a
multi-jet environment such as the LHC, and how these results have
already influenced $e^+e^-$ phenomenology. I will conclude with my
personal view on the challenges that we will have to face at the start
of the Linear Collider (LC), and on what theoretical tools should be
needed to tackle them.

\section{LEP wisdom for LHC Physics}
\label{sec:LEP2LHC}

Before discussing how QCD results from LEP can be exploited at the
LHC, it is worth asking ourselves whether at the LHC precision Physics
has to be limited only to inclusive quantities like $Z$ or $W$
differential cross sections, or can also involve direct measurements
of the hadronic energy-momentum flow. As already stated in the
introduction, for precision purposes it is very difficult to exploit
final-state observables, like event shapes, that are defined in terms
of individual hadrons, since they get huge contributions from poorly
understood phenomena like underlying event or pile-up. However, jets
constructed with modern algorithms are less sensitive to these
effects, and their cross sections can be computed in perturbative QCD
and directly compared to data. In particular, for well separated jets,
fixed order perturbation theory is enough to obtain a reliable
description of data, allowing for measurements of the strong
coupling constant (see for
instance~\cite{Chatrchyan:2011wn}). Furthermore, if the rapidity range
in which jets are observed covers the full detector acceptance,
observables like jet rates become global, and hence can be studied in
the whole range of values of the jet resolution parameters with
all-order resummation techniques~\cite{Banfi:2010xy}. Resummed jet
rates are known from LEP to have small theoretical uncertainties, and
therefore seem the best candidates for precision QCD studies. A close
relative of jet rates is the jet-veto efficiency, for which one can
obtain accurate QCD predictions, which can in turn be exploited in
several New Physics contexts, for instance in Higgs or dark-matter
searches.

Most observables at the LHC however are not suitable for precision
studies, but, like jet masses, are relevant for New Physics
searches. In this case LEP wisdom can be exploited in various ways,
for instance one can try to answer the following questions:
\begin{itemize}
\item Can one reduce contamination from non-perturbative effects in jets?
\item Is there an optimal procedure to filter jets originating from
  boosted object decays?
\item Can we distinguish jets originating from colour singlet decays
  from pure QCD jets?
\end{itemize}
In the following I give examples on how the theoretical methods
developed at LEP have been already exploited to gain some analytical
insight on these issues.

\subsection{Non-perturbative effects in jets}
One of the major theoretical achievements inherited from LEP is
analytical models for hadronisation corrections. Within these
approaches, leading hadronisation corrections to event-shape
distributions and means are given as the product of a perturbatively
calculable coefficient and a single universal non-perturbative
parameter $\alpha_0$, which is extracted from experimental
data~\cite{Dokshitzer:1998pt}. The universality of $\alpha_0$ has been
thoroughly tested at LEP, and is found to hold within
20\%~\cite{Dasgupta:2003iq}. Since analytical hadronisation models
rely basically on the universality of QCD soft radiation, they could
be in principle equally applied to hadronisation corrections in
hadronic collisions. This is what is done for instance in
Ref.~\cite{Dasgupta:2007wa}, where one finds the calculation of the
transverse momentum loss of the leading jet due to hadronisation
$\delta p_{\rm t,had}$, which appears in a variety of jet studies at
the LHC. This quantity is indeed related to the universal parameter
$\alpha_0$, with a coefficient that scales as $1/R$, where $R$ is the
jet radius. Furthermore, since $\delta p_{\rm t,UE}$, the change in
jet $p_t$ due to a hadron background approximately uniform in rapidity
and azimuth (like underlying event or pile-up), is found to scale as
$R^2$, one can compute the radius that minimises the two effects,
which should be then used for precision studies, e.g.\ inclusive jet
transverse momentum spectra. For New Physics searches however, where
one wishes for instance to identify a peak in a jet-mass distribution,
it is also important to minimise the amount of perturbative QCD
radiation that escapes the jet $\delta p_{\rm t,pert}$, which is found
to scale as $\ln (1/R)$.
\label{sec:npjets}
\begin{figure}[htb]
\begin{center}
\resizebox{\textwidth}{!}{
\includegraphics{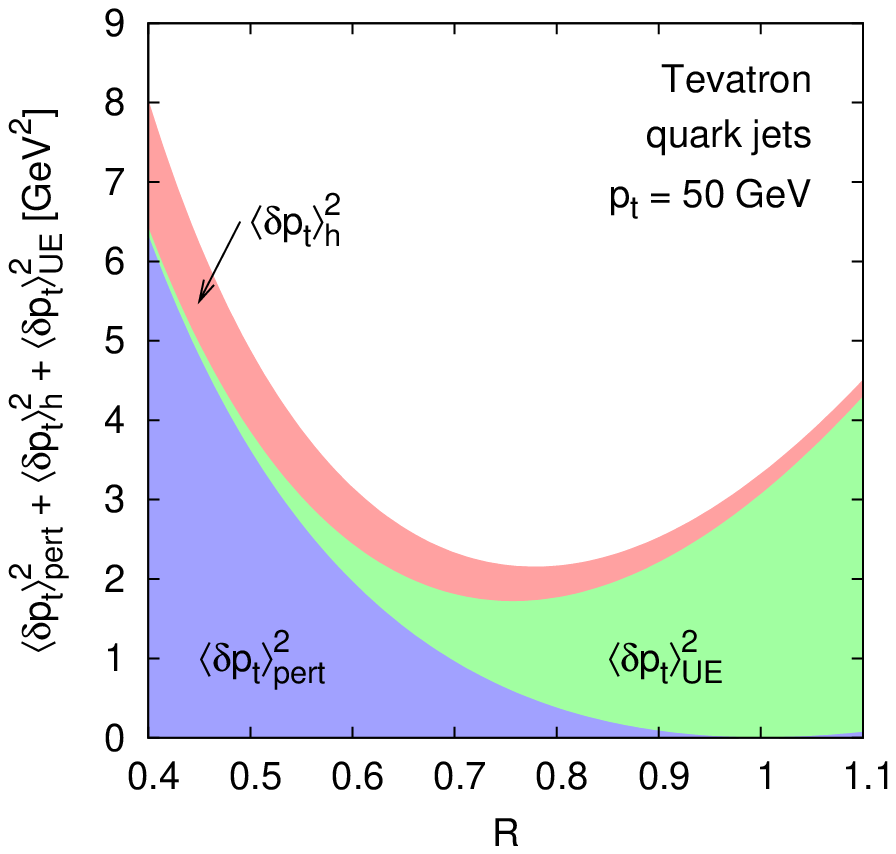}\hspace{2cm}
\includegraphics{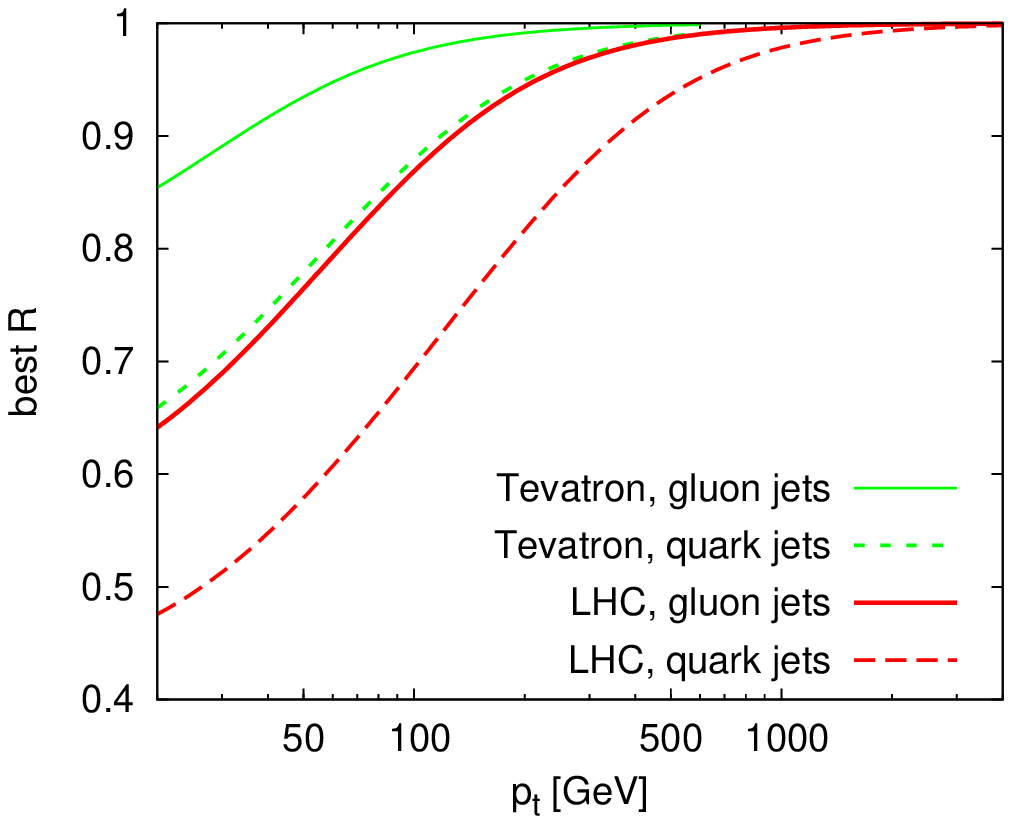}
}
\caption{The average (squared) change in $p_t$ of the leading jet as a
  function of the jet radius (left). The optimal radius given by the
  minimum of $\langle\delta p_t\rangle^2$ as a function of the jet
  $p_t$ for quark and gluon jets at the Tevatron and at the LHC
  (right). Both plots are taken from Ref.~\cite{Dasgupta:2007wa}.
  \vspace{-1.9cm}
}
\label{fig:loss}
\end{center}
\end{figure}
The combined effect of the three sources of $p_t$ loss is illustrated
in Fig.~\ref{fig:loss} (left), from which it is evident that there
exists an optimal radius for which the total $\langle\delta
p_t\rangle^2$ (computed neglecting interference among its different
contributions) is minimised. Since both $\delta p_{\rm t,pert}$ and
$\delta p_{\rm t,had}$ are triggered by QCD radiation, they depend on
the total colour charge of the parton initiating the jet, whilst
$\delta p_{\rm t,UE}$ depends mainly on the centre of mass energy of
the collider. Therefore we expect the optimal radius to change
according to whether we consider quark or gluon jets, and whether we
are at Tevatron or at LHC energies. This is confirmed by actual studies
performed with parton shower event generators, and the resulting
optimal radius as a function of the jet $p_t$ is shown in
Fig.~\ref{fig:loss} (right).

\subsection{Non-global observables and jet filtering}
A relevant topic for New Physics searches at the LHC is the
exploitation of boosted kinematics and jet substructure to detect
high-$p_t$ heavy objects whose decay products fall inside the same jet
(see ~\cite{Altheimer:2012mn} for a recent update). The basic search
strategy consists in clustering each event into ``fat'' jets with a
large radius, and then selecting a candidate jet which should contain
the decay products of the heavy particle one is looking for. The best
known example is a boosted Higgs decaying into a $b\bar b$ pair, where
the candidate Higgs jet must contain at least two separated $b$-tagged
subjets~\cite{Butterworth:2008iy}. Once the candidate jet has been
selected, the problem is how to clean it so that is contains as much
as the Higgs decay products plus QCD radiation originated from them,
and it is least contaminated by initial-state radiation or underlying
event. This is the aim of the filtering procedure, which consists in
reclustering the fat jet with a smaller radius $R_{\rm filt}$ and
reconstructing the candidate Higgs using only the hardest $n_{\rm
  filt}$ subjets. The determination of the best $R_{\rm filt}$ and
$n_{\rm filt}$ relies on the calculation of $\Sigma(\delta M)$, the
fraction of events such that the difference between the Higgs mass and
the jet mass is less than a given $\delta M$. Then one looks for the
value of $\delta M$ for which $\Sigma(\delta M)=f$, with $f$ a given
fraction of events, for instance 68\%: clearly, the smaller $\delta
M$, the better the mass resolution. The quantity $\Sigma(\delta M)$ is
basically an event-shape fraction and, due to the fact that the Higgs
is a colour singlet, can be computed with the theoretical tools
developed for $e^+ e^-$ (non-global) event shapes.
\label{sec:filter}
\begin{figure}[htb]
  \resizebox{\textwidth}{!}{
    \includegraphics[width=.55\textwidth]{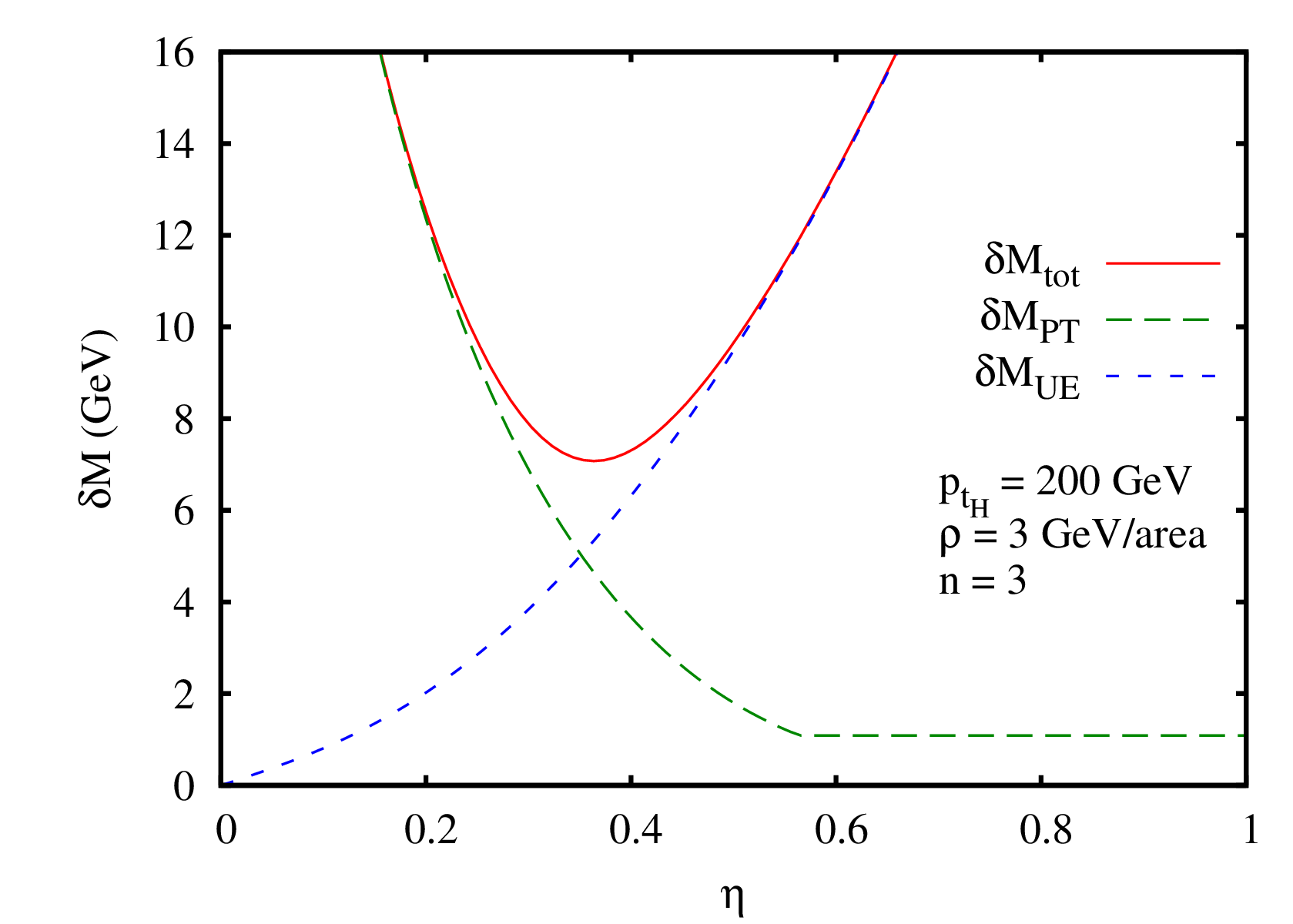}\hspace{2cm}
    \includegraphics[width=.4\textwidth]{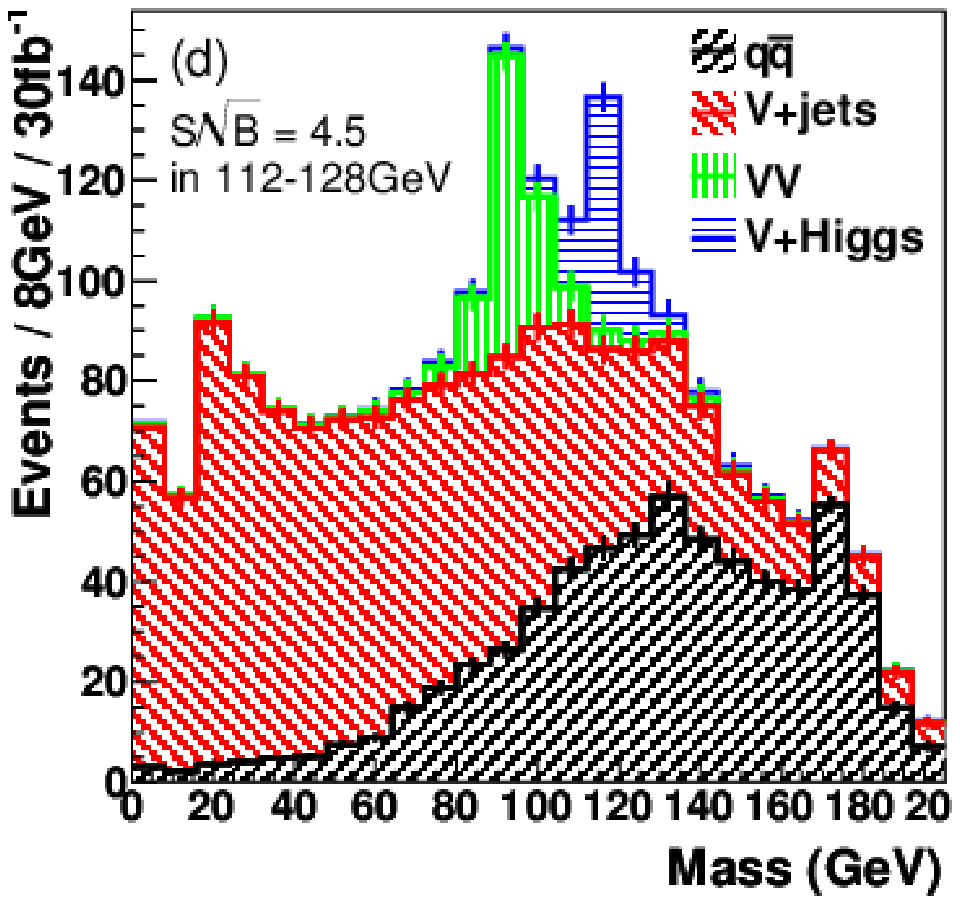}
}
\caption{The width of the Higgs mass peak $\delta M$~\cite{Rubin:2010fc} for
  $n_{\mathrm{filt}}=3$ as a function of $\eta=R_{\rm filt}/R$ (left),
  and the distribution in the invariant mass of the Higgs candidate
  jet corresponding to the selection cuts of
  Ref.~\cite{Butterworth:2008iy} (right).
  \vspace{-1.8cm}
}
\label{fig:Rfilt}
\end{figure}
It is then possible to determine analytically the values of $R_{\rm
  filt}$ and $n_{\rm filt}$ that minimise $\delta M$
(Fig.\ref{fig:Rfilt}, left), and the dedicated study of
Ref.~\cite{Rubin:2010fc} indicates as optimal values $n_{\rm filt}=3$
and $R_{\rm filt} = \min\{R_{b\bar b}/2, 0.3\}$ (where $R_{b\bar b}$ is
the usual $\eta$-$\phi$ distance between the two $b$-subjets). These values
give a good resolution for the Higgs mass peak also after a full event
simulation with parton shower Monte Carlo's (Fig.\ref{fig:Rfilt},
right)~\cite{Butterworth:2008iy}.

\subsection{Colour connections and the ``pull''}
\label{sec:drag}
Many of the heavy objects we wish to observe at the LHC are colour
singlets. This raises the question on whether we can distinguish jets
originating from hadronic decays of colour singlets from pure QCD
jets. Although there is no definitive answer to this question so far,
hints might be gained by studying the QCD radiation pattern in the
interjet region, which is expected to be determined by the colour flow
of each event. Reconstruction of colour connections between jets was
extensively studied at LEP, for instance by counting the number of
hadrons in the interjet region in three-jet events. There one observed
that the hadron multiplicity was different in QCD three-jet events
rather than in $q\bar q\gamma$ events, and the observed difference
could be simply accounted for by considering the colour connections
between the hard emitting partons, the so-called string/drag
effect~\cite{Bartel:1983ij,Aihara:1986hz,Abreu:1995hp,Akers:1995xs}.
An analogous analysis for hadron colliders has been recently
proposed~\cite{Gallicchio:2010sw}. It is based on the so-called
``pull'' vector of a jet, defined as
\begin{equation}
  \label{eq:pull}
  \vec t = \sum_{i \in {\rm jet}} \frac{p_{t,i} |\vec r_i|}{p_{t,jet}} \vec r_i, \qquad
  \vec r_i = (\Delta y_{i,\mathrm{jet}}, \Delta \phi_{i,\mathrm{jet}}) \,. 
\end{equation}
\begin{figure}[htb]
  \centering
  \resizebox{\textwidth}{!}{
\includegraphics[width=.4\textwidth]{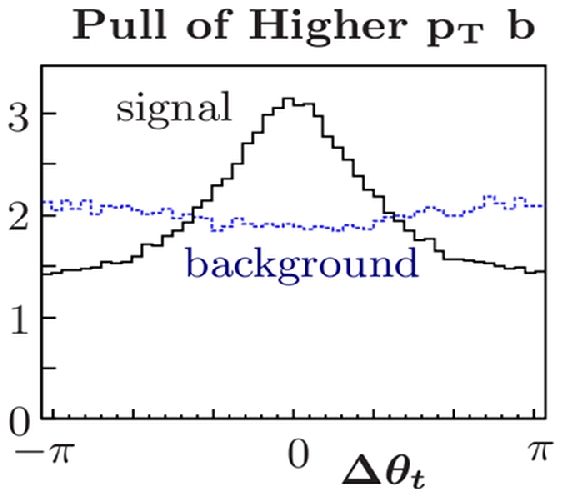}\hspace{1cm}
\includegraphics[width=.5\textwidth]{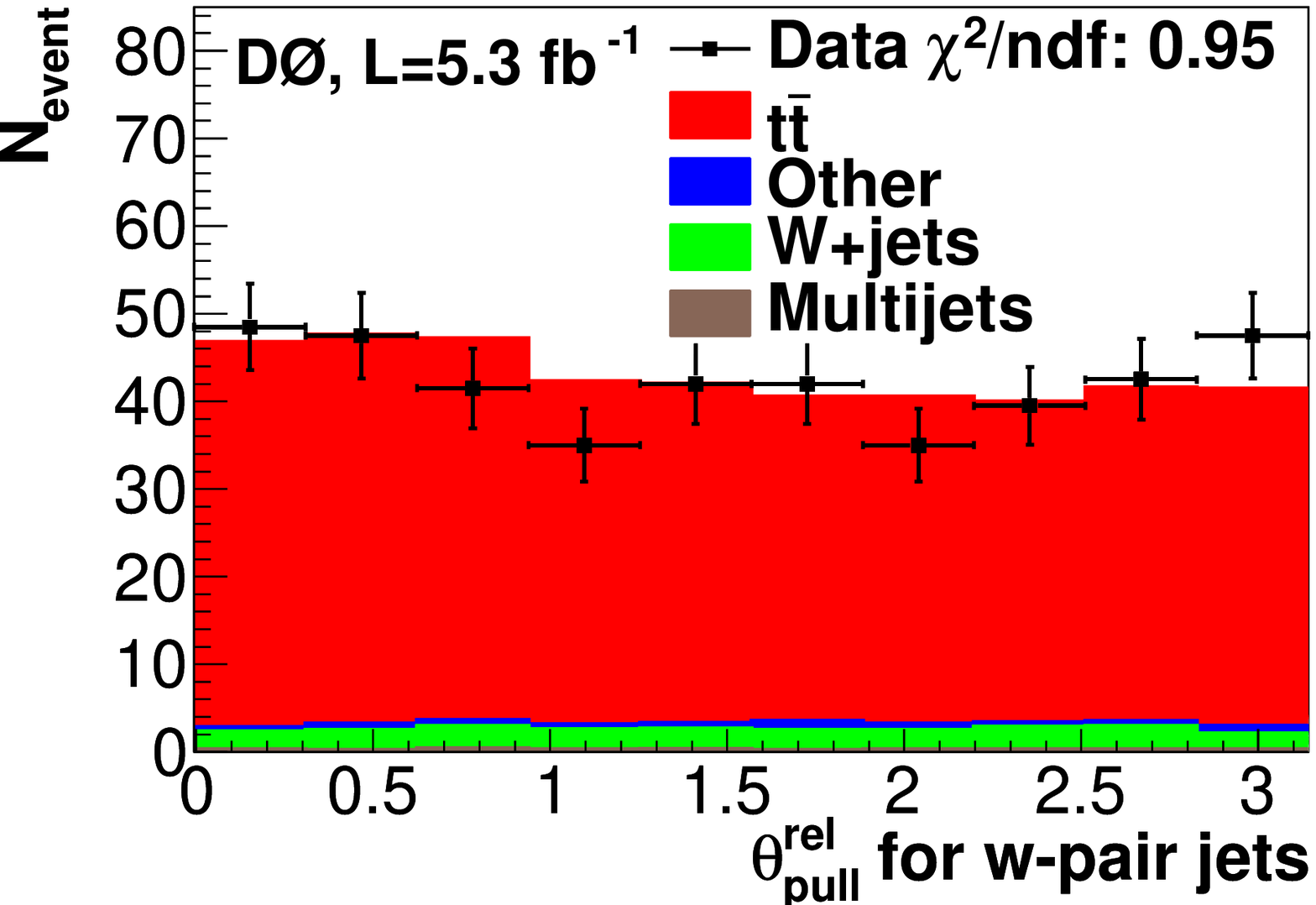}
}
\caption{The distribution in the pull angle $\Delta \theta_t$ for the
  higher $p_t$ $b$-jet in $Z b\bar b$ events, taken from
  Ref.~\cite{Gallicchio:2010sw} (left), and an analogous measurement
  performed at the Tevatron for a pair of jets coming from the decay
  of a $W$ boson in $t \bar t$ events~\cite{Abazov:2011vh} (right).
  \vspace{-1.8cm}
}
  \label{fig:pull}
\end{figure}
The pull distribution ``points'' towards the jet to which the
triggered jet is colour connected.  For instance, following again
Ref.~\cite{Gallicchio:2010sw}, if one considers Higgs production in
association with a $Z$ boson, the distribution in the pull angle
$\Delta\theta_t$ of the higher $p_t$ is peaked around
$\Delta\theta_t=0$, corresponding to the ``position'' of the other
jet, whilst that for the background $Z b \bar b$ is peaked around
$\Delta\theta_t = \pm \pi$, corresponding to the beam (see
Fig.~\ref{fig:pull} left).\footnote{Notice however that a definition
  of the pull vector as in eq.~(\ref{eq:pull}) raises a theoretical
  problem, since at tree level, when a jet consists of a single
  parton, the pull angle is undefined.} Experimental studies in $t\bar
t$ events at the Tevatron confirm this
difference~\cite{Abazov:2011vh}. Indeed, the plot on the right-hand
side of Fig.~\ref{fig:pull} shows the measured distribution in $\Delta
\theta_t$ (labelled $\theta_{\rm pull}^{\rm rel}$ in
Ref.~\cite{Abazov:2011vh}) for any of the two jets coming from the
decay of a $W$ boson. The distribution is peaked around $\theta_{\rm
  pull}^{\rm rel}=0$, corresponding to the ``position'' of the other
jet from $W$ decay. Ref.~\cite{Abazov:2011vh} shows also plots for the
pull angle distribution for the two colour disconnected $b$-jets from
top decay, which is instead peaked towards larger values of
$\theta_{\rm pull}^{\rm rel}$.

\section{QCD predictions for multi-jet events}
\label{sec:LHC2LC}
At LEP the majority of QCD precision studies has been performed for
two-jet events. However, LEP produced many multi-jet
events~\cite{Heister:2003aj,Abdallah:2003xz,Achard:2004sv,Abbiendi:2004qz},
so that at present we have measurements that extend up to the
inclusive six-jet rate~\cite{Heister:2003aj}. However, these multi-jet
events have not been fully exploited for QCD precision studies, the
most notable exception being the three- and four-jet
rates~\cite{Dissertori:2009qa,Dixon:1997th,Nagy:2005gn}. The main
reason for this was the lack of fixed order calculations involving
many legs in the final state. While at LEP one could restrict
experimental analyses to low jet multiplicities, at the LHC many
interesting phenomena, for instance production of top quarks or
supersymmetric particles, involve a large number of jets in the final
state. Notably, already now, there exists data for events $Z$ or $W$
boson production with six additional
jets~\cite{Aad:2012en,Aad:2011qv,Chatrchyan:2011ne}, whose $e^+ e^-$
counterpart is the eight-jet rate! It is therefore clear that one of
the main problem theorists had to face in view of the LHC was how to
perform precision calculations (especially NLO) for multi-leg
processes. The traditional approach based on Feynman diagram looks
prohibitive due to the large number of diagrams (e.g.\ tens of
thousands for processes like $t \bar t b \bar b$, involving four QCD
hard emitters in the final state) which have to be computed. Although,
as the calculation of Ref.~\cite{Denner:2010jp} shows, it is still
possible to perform multi-leg NLO calculations using Feynman diagrams,
in recent years a number of revolutionary ideas changed our way of
looking at one-loop diagrams. The main observation, based on the
pioneering work of Ref.~\cite{Bern:1994zx}, is that the coefficients
of the one-loop master integrals into which any one-loop amplitude can
be decomposed are actually tree-level matrix
elements~\cite{Britto:2004nc,Ossola:2006us,Giele:2008ve}! This was the
starting point of the so-called ``unitarity-cut'' techniques, through
which it is possible to compute one-loop amplitudes as a whole instead
of the individual Feynman diagrams (for a review see
\cite{Ellis:2011cr}, and references therein). With these methods NLO
predictions are nowadays produced at an industrial rate by various
collaborations, such as BLACKHAT~\cite{Ita:2011wn,Bern:2011ep},
HELAC-NLO~\cite{Bevilacqua:2011xh},
ROCKET~\cite{Melia:2010bm,KeithEllis:2009bu},
GOSAM~\cite{Cullen:2011xs}. In the meantime many methods have been
developed to efficiently compute tree-level matrix elements, like
MADGRAPH~\cite{Alwall:2011uj} or COMIX~\cite{Gleisberg:2008fv}. There
are also programs, like ALPGEN~\cite{Mangano:2002ea} or
SHERPA~\cite{Gleisberg:2008ta}, that implement algorithms to
coherently combine tree-level matrix elements to parton showers. Last
but not least, in recent years new methods have been developed to
match even NLO calculations to parton
showers~\cite{Frixione:2002ik,Nason:2004rx}, nowadays automated in the
aMC@NLO~\cite{Frederix:2011ss} and POWHEG-BOX~\cite{Alioli:2010xd}
frameworks.

\begin{figure}[htb]
  \centering
  \resizebox{\textwidth}{!}{
    \includegraphics[height=.5\textwidth,angle=270]{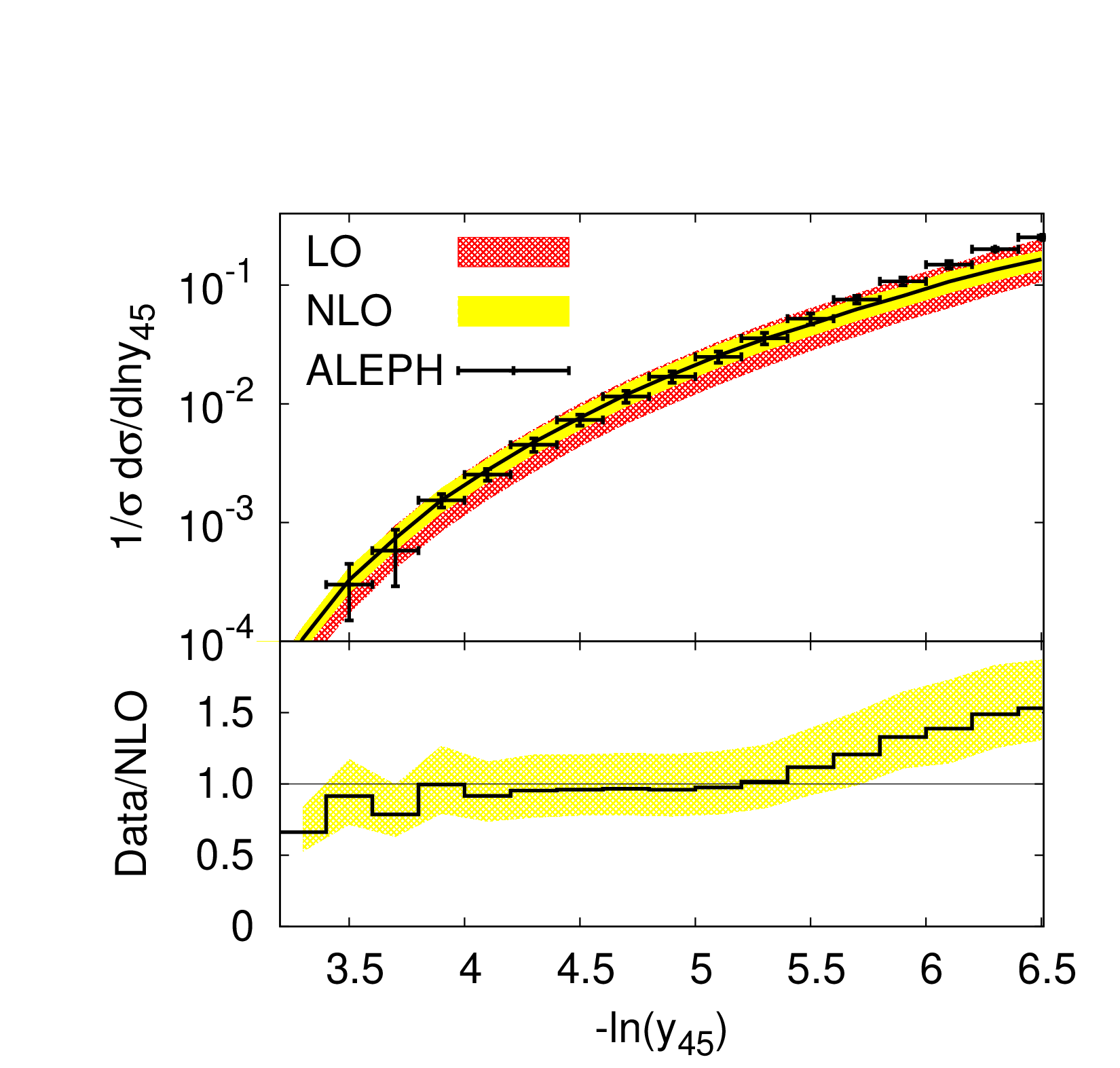}
    \includegraphics[height=.5\textwidth,angle=270]{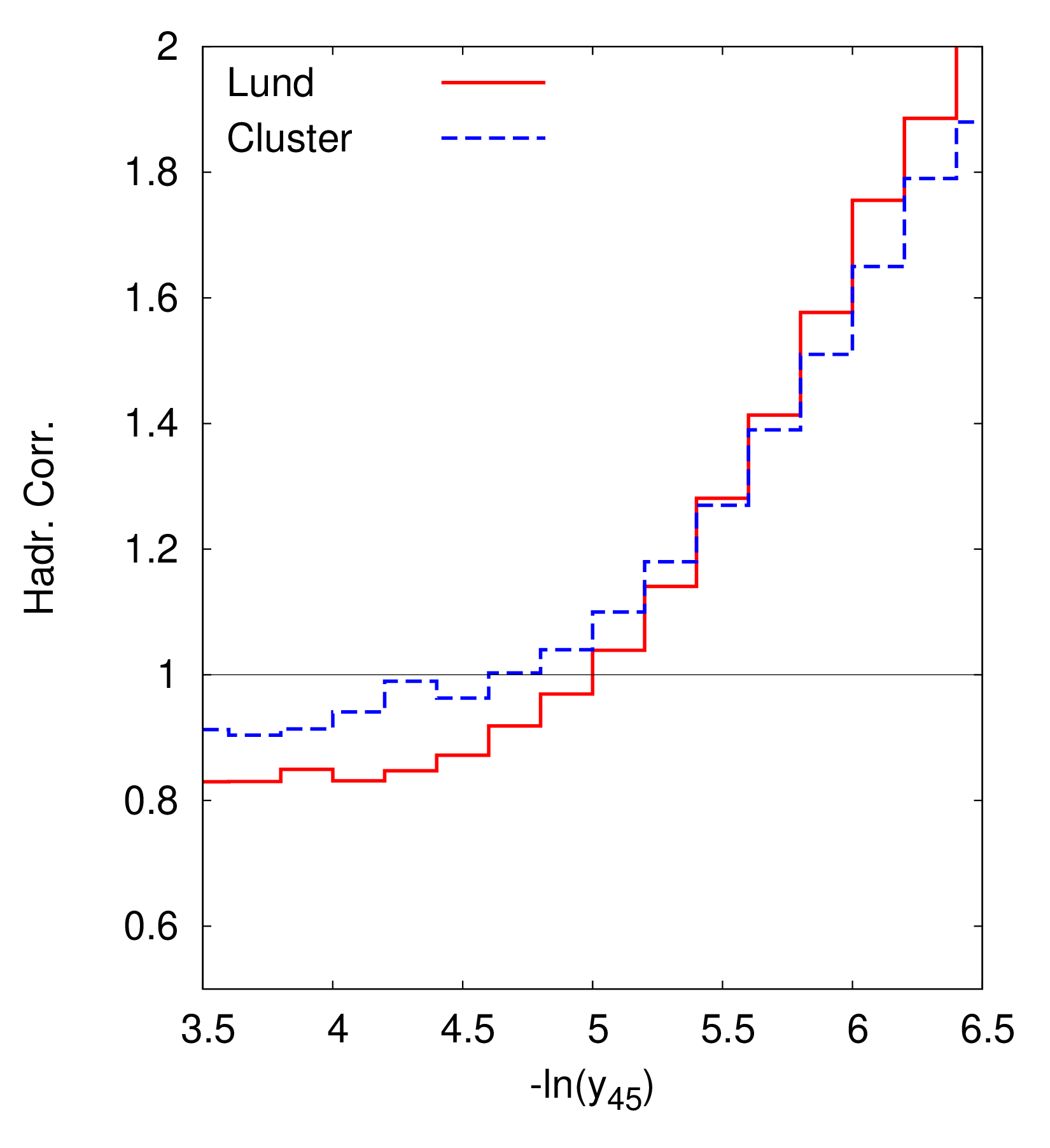}
}
\caption{The NLO distribution in (log of) the 5-jet resolution
  $y_{45}$ compared to ALEPH data (left), and the corresponding
  hadronisation corrections obtained with SHERPA (right). Both plots
  are taken from Ref.~\cite{Frederix:2010ne}.  
  \vspace{-1.8cm} }
  \label{fig:y45}
\end{figure}

Such enormous progress had consequences also for $e^+ e^-$ precision
studies. For instance, for the first time it was possible to tackle
the NLO calculation of the five-jet rate by crossing matrix elements
used for $Z$ plus three jets at NLO~\cite{Frederix:2010ne}. The
resulting theoretical analysis, in particular the extraction of a
value of $\alpha_s(M_Z)$, took also advantage of the matching between
tree-level five-jet matrix elements and parton shower implemented in
the SHERPA Monte Carlo. Indeed, only using SHERPA was it possible to
obtain a reliable estimate of hadronisation corrections, and hence a
precise measurement of $\alpha_s(M_Z)$ (see Fig.~\ref{fig:y45}). The
limit on the jet multiplicity is nowadays being pushed further and
further, and at the moment there exist (leading colour) NLO
calculations for $e^+e^-$ up to seven jets~\cite{Becker:2011vg}. It
would be great to compare these predictions to LEP data, so as to have
consistent extractions of $\alpha_s(M_Z)$ from all measured jet rates.

\section{Outlook}
\label{sec:fine}
The NLO revolution is just one example of the theoretical progress
that has been triggered by LHC Physics in recent years. For processes
like the production of colour singlets NNLO calculations are already
available~\cite{Catani:2011qz,Ferrera:2011bk,Catani:2009sm,Catani:2007vq,Anastasiou:2004xq,Anastasiou:2003yy}
while considerable progress has been made towards NNLO predictions for
top-antitop~\cite{Czakon:2011ve} or dijet
production~\cite{GehrmannDeRidder:2011aa}. Given the complexity of
two-loop calculations, many people have also tried to investigate
whether the structure of QCD amplitudes could be deduced from general
principles rather than obtained only through explicit
calculations. This research stream involved on one hand the use of
factorisation properties of gauge theories to arrive at a general
formula for the infrared structure of gauge
theories~\cite{Gardi:2009qi,Becher:2009qa}. On the other hand, also
hard non singular contributions were investigated in theories with a
high degree of symmetry, like $\mathcal{N}=4$ Super Yang-Mills, hoping
to be able to solve them at the quantum level (see for
instance~\cite{Beisert:2010jr}). The latter studies have lead to the
discovery of the simpler representation of multi-loop amplitudes in
terms of mathematical objects called
symbols~\cite{Goncharov:2010jf}. The hope is to be able to associate
to each amplitude its symbol content, so as to avoid completely the
explicit calculation of loop integrals.

Given the theoretical advances I have described so far, how can we
imagine the state of the art of precision calculations at the start of
the linear collider? Let us consider for instance the Higgsstrahlung
process, the most widely used for Higgs searches at LEP, with both the
Higgs and the recoiling vector boson decaying hadronically. The
theoretical description of both signal and backgrounds (e.g.\ $e^+
e^-$ to four jets) will be very different from that of LEP
days. Definitely higher order corrections, both QCD and electro-weak,
will be available at NNLO, and probably we will know the all-order
structure of the dominant virtual corrections. Sophisticated methods
based on jet substructure will be able to discriminate the signal from
the backgrounds. Experimental analyses will also take advantage of the
fact that all parton shower event generators will be matched to NLO
matrix elements.

Concerning precision Physics, for two-jet event shapes hadronisation
corrections will be very small, so that, just using the already
available NNLO+NNLL predictions, we could have a measurement of
$\alpha_s(M_Z)$ at the permille accuracy.  Jet rates had already very
small hadronisation corrections at LEP, at LC they will have basically
none. In this case we already have NNLO predictions for three-jet
production, and probably we will have them for four-jet production as
well. Unfortunately no resummation beyond NLL accuracy is available
for jet rates so far. An improvement in this direction would probably
allow for the most precise determination of $\alpha_s(M_Z)$ ever.

I would like to conclude with a couple of remarks on non-perturbative
effects, which are in fact the everlasting unknown in all collider
experiments. The LC will not only be a precision machine, but also a
means of investigation of those effects, especially in multi-jet
events. For instance, when considering three-jet events, due to the
extra radiation from a gluon, leading hadronisation corrections are
expected to be roughly twice as large as in two-jet events. This
feature, at LEP energies, made them too large to be allowed to neglect
the contribution of subleading corrections, as was done for two-jet
events.  At the LC instead, non-perturbative corrections to three-jet
event shapes like the $D$-parameter are of the same order of magnitude
of the corresponding ones to two-jet event shapes at LEP. Therefore,
more studies of the universality of the non-perturbative parameter
$\alpha_0$ could be performed, opening for the first time the
possibility of making quantitative statements about hadronisation from
a gluon in a multi-jet environment. Last but not least, there might be
experimental high-luminosity setups for the LC which imply
contamination of signal events from pile-up. We hope that the LHC will
teach us how to model this effect better and better, so as to be able
to properly deal with it before the start of the LC.

\section*{Acknowledgements}
I am deeply grateful to the organisers of the workshop for the
invitation and for the pleasant and stimulating atmosphere they were
able to create. I also would like to thank Stephen Kluth, Gavin Salam
and Giulia Zanderighi for helpful comments and suggestions during the
preparation of my talk.

 \end{document}